\documentclass[aps,prl,twocolumn,superscriptaddress]{revtex4}
\usepackage{mathrsfs}
\usepackage{amssymb, amsbsy, amsmath, latexsym, dsfont, array, layout,
graphicx,mathrsfs,color,ulem,bm}
\usepackage{amsfonts}
\usepackage{epsfig}
\usepackage{graphics}
\usepackage{bbold}
\usepackage{psfrag}
\usepackage{mathcomp}
\usepackage{subfigure}
\usepackage{verbatim}
\usepackage[colorlinks,citecolor=blue]{hyperref}

\begin{document}

\title{Fixed points and emergent topological phenomena in a parity-time-symmetric quantum quench}
\author{Xingze Qiu}
\affiliation{Key Laboratory of Quantum Information, University of Science and Technology
of China, CAS, Hefei, Anhui, 230026, China}
\affiliation{Synergetic Innovation Center of Quantum Information and Quantum Physics,
University of Science and Technology of China, Hefei, Anhui 230026, China}
\author{Tian-Shu Deng}
\affiliation{Key Laboratory of Quantum Information, University of Science and Technology
of China, CAS, Hefei, Anhui, 230026, China}
\affiliation{Synergetic Innovation Center of Quantum Information and Quantum Physics,
University of Science and Technology of China, Hefei, Anhui 230026, China}
\author{Ying Hu}\email{huying@sxu.edu.cn}
\affiliation{
State Key Laboratory of Quantum Optics and Quantum Optics Devices, Institute of Laser Spectroscopy, Shanxi University, Taiyuan, Shanxi 030006, China}
\affiliation{Collaborative Innovation Center of Extreme Optics, Shanxi University, Taiyuan, Shanxi 030006, China}
\author{Peng Xue}\email{gnep.eux@gmail.com}
\affiliation{Beijing Computational Science Research Center, Beijing 100084, China}
\affiliation{Department of Physics, Southeast University, Nanjing 211189, China}
\affiliation{State Key Laboratory of Precision Spectroscopy, East China Normal University, Shanghai 200062, China}
\author{Wei Yi}
\email{wyiz@ustc.edu.cn}
\affiliation{Key Laboratory of Quantum Information, University of Science and Technology
of China, CAS, Hefei, Anhui, 230026, China}
\affiliation{Synergetic Innovation Center of Quantum Information and Quantum Physics,
University of Science and Technology of China, Hefei, Anhui 230026, China}

\begin{abstract}
We identify emergent topological phenomena such as dynamic Chern numbers and dynamic quantum phase transitions in quantum quenches of the non-Hermitian Su-Schrieffer-Heeger Hamiltonian with parity-time ($\mathcal{PT}$) symmetry.
Their occurrence in the non-unitary dynamics are intimately connected with fixed points in the Brillouin zone, where the states do not evolve in time. We construct a theoretical formalism for characterizing topological properties in non-unitary dynamics within the framework of biorthogonal quantum mechanics, and prove the existence of fixed points for quenches between distinct static topological phases in the $\mathcal{PT}$-symmetry-preserving regime. We then reveal the interesting relation between different dynamic topological phenomena through the momentum-time spin texture characterizing the dynamic process. For quenches involving Hamiltonians in the $\mathcal{PT}$-symmetry-broken regime, these topological phenomena are not ensured.
\end{abstract}

\maketitle

The exploration of topological matter constitutes a major theme in modern physics~\cite{HKrmp10,QZrmp11}. With rapid progress in the discovery and understanding of topological phases in solid-state materials, a challenging quest lies in extending the study of conventional topological matter to novel regimes. Prominent examples include the investigation of emergent topological properties in out-of-equilibrium dynamics~\cite{Demler10,Zoller11,Zollerdiss,Levin13,Gogolin15,rigol,Bhaseen15,Dora15,Lindner16,Vishwanath16,Sondhi16,Heyl13,Heyl15,Heyl17,BH16,Balatsky,Bhaseen16,Budich16,Refael16,Sondhi17,Zhai17,Chen17,Ueda17,Chang18, Xiong-Jun1802, Xiong-Jun1807} and the characterization of topological phases in non-Hermitian systems~\cite{RL09,ESHK11,LH13,schomerus2013,RLL16,KMKO16, Gong17,Lieu18,UedaNH,UedaPT,Wang1803,Wang1804,Kunst18}. With the flexible controls afforded by synthetic systems such as ultracold atoms and engineered photonic configurations, the experimental implementation of these interesting scenarios is already within reach~\cite{ETHcoldatom14,Weitenberg2016,Jo17,Weitenberg1709,Weitenberg17,Bellecnc,Weimannnm,Zeunerprl,PTsymm2,pxprl}.

An exemplary situation for the study of topological properties in out-of-equilibrium dynamics is the quantum quench of a topological system,
where the ground state of the initial Hamiltonian $H^{\text{i}}$ is subject to a unitary time evolution governed by the final Hamiltonian $H^{\text{f}}$. Whereas the topological invariant characterizing the instantaneous state is unchanged during the unitary dynamics~\cite{Bhaseen15,rigol}, previous studies have revealed the emergence of intriguing phenomena such as dynamic quantum phase transitions (DQPTs)~\cite{BH16, Balatsky,Dora15,Heyl13,Heyl15,Heyl17,Weitenberg17,iondtop} and quantized non-equilibrium Hall responses in quench processes~\cite{Budich16,Refael16,Bhaseen16}. Further, in a series of recent theoretical and experimental studies, it has been established that dynamic topological invariants can be defined in unitary quantum quenches, which are related to the topology of initial and final Hamiltonians in equilibrium~\cite{Zhai17,Chen17,Ueda17,Weitenberg1709}.

Here arises an interesting question: what if the quench dynamics is non-unitary and governed by non-Hermitian Hamiltonians? The question is particularly relevant in light of recent studies on topological phenomena in parity-time($\mathcal{PT}$)-symmetric non-Hermitian systems~\cite{schomerus2013,KMKO16,Bellecnc,Weimannnm,PTsymm2,UedaPT}.
Under $\mathcal{PT}$ symmetry, eigenenergies of a non-Hermitan Hamiltonian are entirely real in the $\mathcal{PT}$-symmetry-preserving regime,
in contrast to regimes with spontaneously broken $\mathcal{PT}$ symmetry~\cite{BB98,BBJ02,B07}. Whereas it has been shown that $\mathcal{PT}$ symmetry has profound impact on topological properties of static topological phases~\cite{schomerus2013,KMKO16,Bellecnc,Weimannnm,PTsymm2,UedaPT}, its role in emergent topological phenomena in dynamic processes has never been explored. A particular difficulty lies in the theoretical description of dynamics generated by non-Hermitian Hamiltonians, which is non-unitary even when the system is in the $\mathcal{PT}$-symmetry-preserving regime with real eigen spectra. Searching for topological phenomena in dynamics of non-Hermitian systems thus requires theories beyond the conventional paradigm in Hermitian systems under unitary time evolution~\cite{rigol,Refael16, Zhai17,Chen17,Ueda17,Weitenberg1709}.

In this work, we study the quench dynamics of a non-Hermitian Su-Schrieffer-Heeger (SSH) Hamiltonian with $\mathcal{PT}$ symmetry. We construct a novel theoretical formalism based on biorthogonal quantum mechanics, which is capable of characterizing dynamic topological properties in non-Hermitian systems. We then demonstrate how fixed points in the non-unitary time evolution give rise to dynamic Chern numbers and DQPTs, when the system is quenched across topological phase boundaries in the $\mathcal{PT}$-symmetry-preserving regime. Interestingly, while dynamic Chern numbers derive from emergent skyrmions in the momentum-time-space spin texture describing the dynamic process, critical points of DQPTs manifest as vortex cores in the same spin texture.
By contrast, if either the pre- or post-quench Hamiltonian is in the $\mathcal{PT}$-symmetry-broken regime, the appearance of fixed points are not guaranteed, and the aforementioned topological phenomena are typically absent.

\begin{figure}
\includegraphics[width=8.5cm]{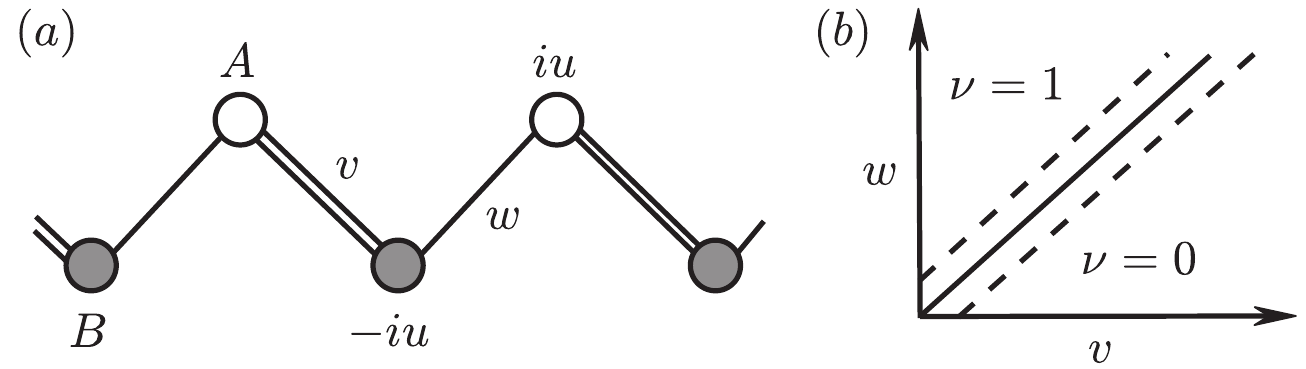}
\caption{(a) Schematic illustration of the non-Hermitian SSH Hamiltonian. (b) Topological phase diagram. The solid line is the topological phase boundary with $w=v$. Dashed lines are the boundaries between $\mathcal{PT}$-symmetry-preserving and broken regimes with $|w-v|=u$.}
\label{fig:SSH}
\end{figure}

{\it $\mathcal{PT}$-symmetric SSH model:---}
As illustrated in Fig.~\ref{fig:SSH}(a), we consider the non-Hermitian SSH model with alternating gain and loss on adjacent sites under the periodic boundary condition~\cite{SSH,SSHPT}
\begin{align}
H=\sum_{j=1}^{L-1} \left(v a^{\dag}_j b_{j} + w a^{\dag}_{j+1} b_j+\text{H.c.}\right)+iu \sum_{j=1}^L  (a^{\dag}_ja_j-b^{\dag}_jb_j),\label{eqn:H}
\end{align}
where $a^{\dag}_j$ ($b^{\dag}_j$) is the creation operator on $A$ ($B$) sub-lattice at site $j$, $L$ is the total number of unit cells, the tunneling amplitudes $v,w$ and the gain-loss rate $u$ satisfy $u,v,w\geq 0$, and $\text{H.c.}$ stands for Hermitian conjugation.

Hamiltonian (\ref{eqn:H}) possess $\mathcal{PT}$ symmetry, as $\mathcal{PT} H \left(\mathcal{PT}\right)^{-1}=H$, with the parity operator $\mathcal{P}a(b)_j\mathcal{P}=b(a)_{L+1-j}$, and the time reversal operator $\mathcal{T}i\mathcal{T}^{-1}=-i$, respectively. Under $\mathcal{PT}$ symmetry, the eigenspectrum of (\ref{eqn:H}) is entirely real if all eigenstates are simultaneous eigenstates of the $\mathcal{PT}$-symmetry operator. In this case, the Hamiltonian is in the $\mathcal{PT}$-symmetry-preserving regime. Otherwise, the Hamiltonian is in the $\mathcal{PT}$-symmetry-broken regime, where some eigenstates spontaneously break $\mathcal{PT}$ symmetry and acquire imaginary eigenergies.
The transition between the $\mathcal{PT}$-symmetry-preserving and broken regimes can be derived
by examining the Bloch Hamiltonian $H_k=\bm{h}(k)\cdot\boldsymbol{\sigma}$ at momentum $k$, where $\bm{\sigma}=(\sigma_1,\sigma_2,\sigma_3)$ and $\sigma_\alpha$ ($\alpha=1,2,3$) are the Pauli matrices.
The complex vector $\bm{h}(k)=(h_1,h_2,h_3)$, with $h_1=w\cos k+v$, $h_2=w\sin k$, and $h_3=iu$. As the eigenenergy is given by $\epsilon_{\mu}=\mu E_k$ ($\mu=\pm$) with $E_k=\sqrt{w^2+v^2+2wv\cos k-u^2}$, the Hamiltonian is in the $\mathcal{PT}$-symmetry-preserving regime when $u<|v-w|$, where $\epsilon_{\pm}$ is real for all $k$.

The $\mathcal{PT}$-symmetric SSH Hamiltonian possesses topological properties, guaranteed by the so-called pseudo-anti-Hermicity with $\eta H^{\dag}\eta^{-1}=-H$~\cite{ESHK11}, where $\eta=\sigma_z$.
As a result, $\mathcal{PT}$-symmetry-broken topological edge states with purely imaginary eigenenergies emerge at the boundary between bulks of different topological phases.
Topologically inequivalent phases can be distinguished by the generalized winding number $\nu=\varphi_\text{B}/2\pi$. Here, the global Berry phase $\varphi_{\text{B}}$ is~\cite{LH13,Lieu18,GW88,supp}
\begin{align}
\varphi_B=-i\sum_{\mu=\pm}\oint dk\frac{\langle\chi_{\mu}|\frac{\partial}{\partial k}|\psi_{\mu}\rangle}{\langle\chi_{\mu}|\psi_{\mu}\rangle},\label{eqn:gb}
\end{align}
where the integral runs over the first Brillioun zone (1BZ), and the right (left) eigenvector is defined as $H_k|\psi_\mu\rangle=\epsilon_\mu|\psi_\mu\rangle$ ($H_k^{\dag}|\chi_{\mu}\rangle=\epsilon^*_{\mu}|\chi_{\mu}\rangle$).
When $v<w$, the system is topologically non-trivial, as $\nu=1$. In contrast, when $v>w$, we have $\nu=0$ and the system is topologically trivial.
In Fig.~\ref{fig:SSH}(b), we show the topological phase diagram as well as the boundary between $\mathcal{PT}$-symmetry-preserving and broken regimes. Notably, the $\mathcal{PT}$-symmetry-broken regime lies in the vicinity of the topological phase boundary.

Below we consider quantum quenches of Hamiltonian (\ref{eqn:H}), where the parameters $(u,v,w)$ undergo abrupt changes at $t=0$.
Due to the lattice translational symmetry of (\ref{eqn:H}), the dynamics in different $k$-sectors can be analyzed independently.

{\it Visualizing non-unitary dynamics on the Bloch sphere:---}
To account for the non-unitary dynamics of quenched non-Hermitian systems, we invoke the notion of biorthogonal quantum mechanics~\cite{DCB}. Denoting the initial Hamiltonian by $H^\text{i}$, the initial state in each $k$-sector $|\psi_-^{i}\rangle$, with $H^{\text{i}}_k|\psi_-^{i}\rangle=\epsilon^\text{i}_-|\psi_-^{i}\rangle$, evolves under the final Hamiltonian $H^{\text{f}}_k$ according to $|\psi_k(t)\rangle=\sum_{\mu=\pm} c_\mu e^{-i\epsilon^{\text{f}}_\mu t}|\psi^{\text{f}}_\mu\rangle$, where $\epsilon_\mu^{\text{f}}=\mu E_k^{\text{f}}$ is the eigenenergy of $H_k^{\text{f}}$. We have defined $c_\mu=\langle\chi^{\text{f}}_\mu|\psi^{\text{i}}_-\rangle$, where $\langle\chi^{\text{f}}_\mu|$ $\left(|\psi^{\text{f}}_\mu\rangle\right)$ is the left (right) eigenvector of $H_k^{\text{f}}$, with the biorthonormal conditions $\langle\chi^{\text{f}}_\mu|\psi^{\text{f}}_\nu\rangle=\delta_{\mu\nu}$ and $\sum_\mu|\psi^{\text{f}}_\mu\rangle\langle\chi^{\text{f}}_\mu|=1$.

To characterize the non-unitary time evolution in the biorthogonal basis $\left\{|\psi^{\text{f}}_\mu\rangle,|\chi^{\text{f}}_\mu\rangle\right\}$, we define an associated state of $|\psi_k(t)\rangle$ as $\langle\chi_k(t)|=\sum_\mu c^*_\mu e^{i\epsilon^{\text{f}*}_\mu t}\langle\chi^{\text{f}}_\mu|$~\cite{DCB}, with the normalization condition $\langle\chi_k(0)|\psi_k(0)\rangle=1$. The non-unitary time evolution of the system is then captured by the non-Hermitian density matrix $\rho(k,t)=\frac{|\psi_k(t)\rangle\langle\chi_k(t)|}{\langle\chi_k(t)|\psi_k(t)\rangle}$, such that the expectation value of any operator $A_k$ is expressed as $\text{Tr} (\rho A_k)$~\cite{DCB}.
We further write
\begin{align}
\rho(k,t)=\frac{1}{2}\left[\tau_0+\bm{d}(k,t)\cdot\boldsymbol{\tau}\right],
\label{Eq:densitymatrix}
\end{align}
where $\bm{d}(k,t)=(d_1,d_2,d_3)$, $\bm{\tau}=(\tau_1,\tau_2,\tau_3)$, and we have defined the matrices
$\tau_\gamma=\sum_{\mu,\nu=\pm}|\psi_{\mu}^{\text{f}}\rangle\sigma^{\mu\nu}_\gamma\langle\chi_{\nu}^{\text{f}}|$ ($\gamma=0,1,2,3$). Here $\sigma^{\mu\nu}_{\gamma}$ is the matrix element of $\sigma_{\gamma}$, and $\sigma_0$ is the $2\times 2$ identity matrix. Note that $\{\tau_\beta\}$ fulfil the standard $\mathfrak{su}(2)$ commutation relations.

As a key ingredient of our theoretical construction, the introduction of $\{\tau_\beta\}$ matrices makes the corresponding $\bm{d}(k,t)$ a real unit vector on the Bloch sphere $S^2$, even as the dynamics is non-unitary. Equation (\ref{Eq:densitymatrix}) thus allows a geometrical picture for understanding non-unitary dynamics and forms the basis for subsequent characterization of dynamic topological properties therein.

\begin{figure}
\includegraphics[width=8cm]{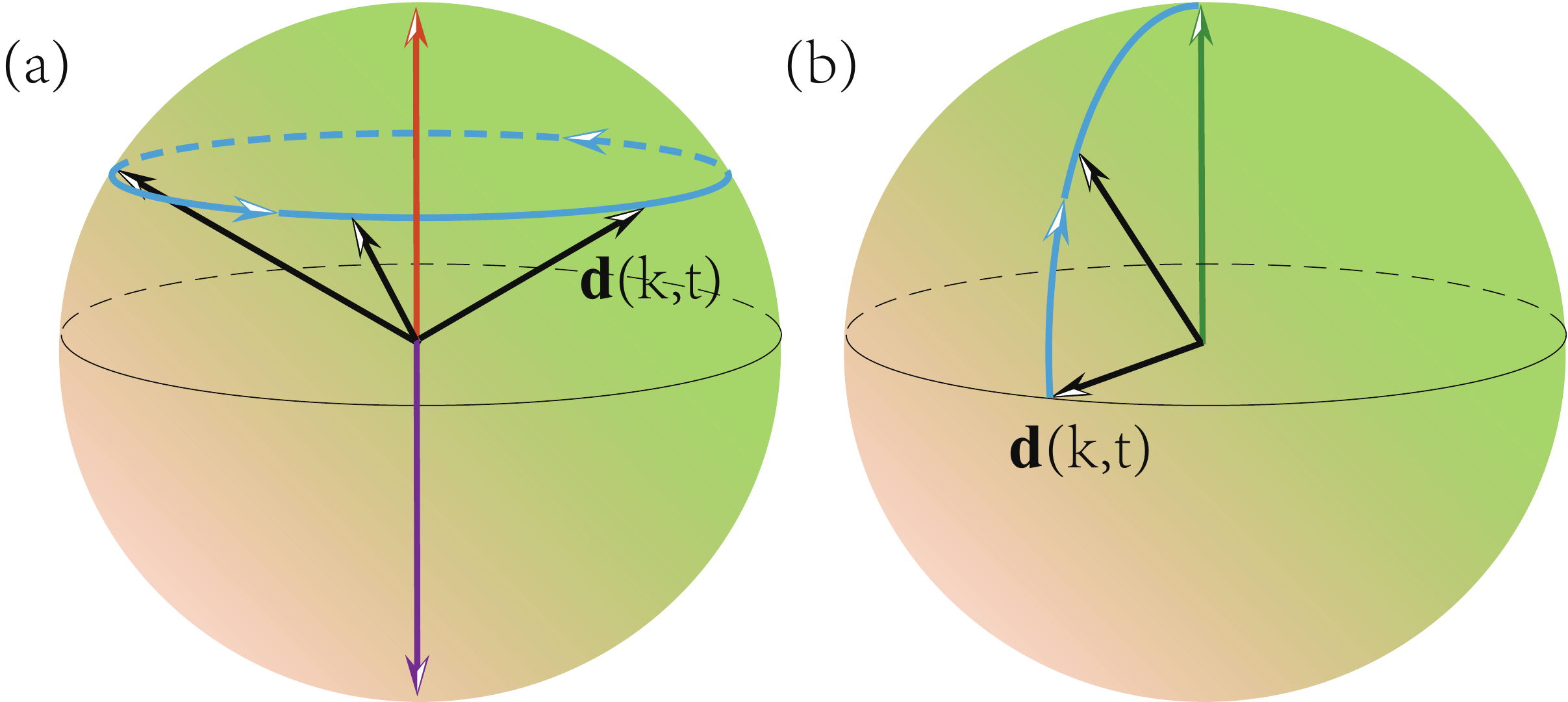}
\caption{Schematic illustrations of the time evolution of $\bm{d}(k,t)$ (black) on the Bloch sphere for the $k$-sector with: (a) real $E^\text{f}_k$, and (b) imaginary $E^\text{f}_k$ (assuming $\text{Im}(E_{k}^{\text{f}})>0$). The orange and purple vectors correspond, respectively, to fixed points with $c_-=0$ and $c_+=0$. The green vector in (b) indicates the long-time steady state.
}
\label{fig:sphere}
\end{figure}

{\it Fixed points in non-unitary dynamics:---}
When $E_{k}^{\text{f}}$ is real, $\bm{d}(k,t)$ and hence the density matrix $\rho(k,t)$ are oscillatory in time, with a momentum-dependent period $t_0=\pi/E^\text{f}_k$~\cite{supp}. This corresponds to a periodic rotation of $\bm{d}(k,t)$ around the poles of the Bloch sphere, as illustrated in Fig.~\ref{fig:sphere}(a). Importantly, when $\bm{d}(k,0)$ is on the poles of the Bloch sphere, the density matrix becomes time independent. This occurs at momenta $k_m$ with either $c_-(k_m)=0$ (north pole) or $c_+(k_m)=0$ (south pole), which are identified as two different kinds of fixed points.
In contrast, when $E_{k}^{\text{f}}$ is imaginary, $\bm{d}(k,t)$ always starts from the equator at $t=0$ and approaches the north pole in the long-time limit [Fig.~\ref{fig:sphere}(b)], i.e., $\rho(k,t)$ exponentially approaches a steady-state value~\cite{supp}. In this case, there are no fixed points in the dynamics.

Based on the understanding above, it is straightforward to show that the number of fixed points with $c_+=0$ or $c_-=0$ should be at least $\left|\nu^{\text{i}}-\nu^{\text{f}}\right|$ each, provided both $H^\text{i}$ and $H^\text{f}$ belong to the $\mathcal{PT}$-symmetry-preserving regime with completely real eigenspectra~\cite{supp}. Here $\nu^\beta$ ($\beta=\text{i},\text{f}$) are the generalized winding numbers of $H^\beta$.
On the other hand, when $H^\text{f}$ is in the $\mathcal{PT}$-symmetry-broken regime, the corresponding $E^{\text{f}}_k$ becomes imaginary for a certain range of $k$, and the existence of fixed points are no longer guaranteed.

In the following, we mainly focus on the case where both $H^\text{i}$ and $H^\text{f}$ are in the $\mathcal{PT}$-symmetry-preserving regime. When the system is quenched across the topological phase boundary, fixed points divide the BZ into a series of submanifolds, where the states at two ends of each given submanifold do not evolve in time. This feature gives rise to dynamic topological phenomena as we will detail in the remainder of the work.

{\it Dynamic Chern number:---}
When $H^\text{f}$ is in the $\mathcal{PT}$-symmetry-preserving regime, the periodic oscillation of the density-matrix evolution gives rise to an $S^1$ topology in the time evolution.
In the presence of fixed points, each submanifold between two adjacent fixed points can be combined with the $S^1$ topology in time to form a momentum-time manifold $S^2$, which can be mapped to the Bloch sphere associated with the vector $\bm{d}(k,t)$~\cite{Chen17,Ueda17}.
These $S^2\rightarrow S^2$ mappings define a series of dynamic Chern numbers
\begin{align}
C_{mn} = \frac{1}{4\pi}\int_{k_m}^{k_{n}}dk\int_{0}^{t_0}dt[\bm{d}(k,t)\times\partial_t \bm{d}(k,t)]\cdot\partial_k \bm{d}(k,t),
\end{align}
where $k_m$ and $k_n$ denotes two neighboring fixed points.
For quenches between Hamiltonians with different winding numbers, the dynamic Chern numbers are quantized, with values dependent on the nature of fixed points at $k_m$ and $k_{n}$~\cite{supp}: $C_{mn}=1$ when $c_+(k_{m})=0$ and $c_-(k_{n})=0$; $C_{mn}=-1$ when $c_-(k_{m})=0$ and $c_+(k_{n})=0$. When the two fixed points are of the same kind, $C_{mn}=0$.

\begin{figure}
\includegraphics[width=9cm]{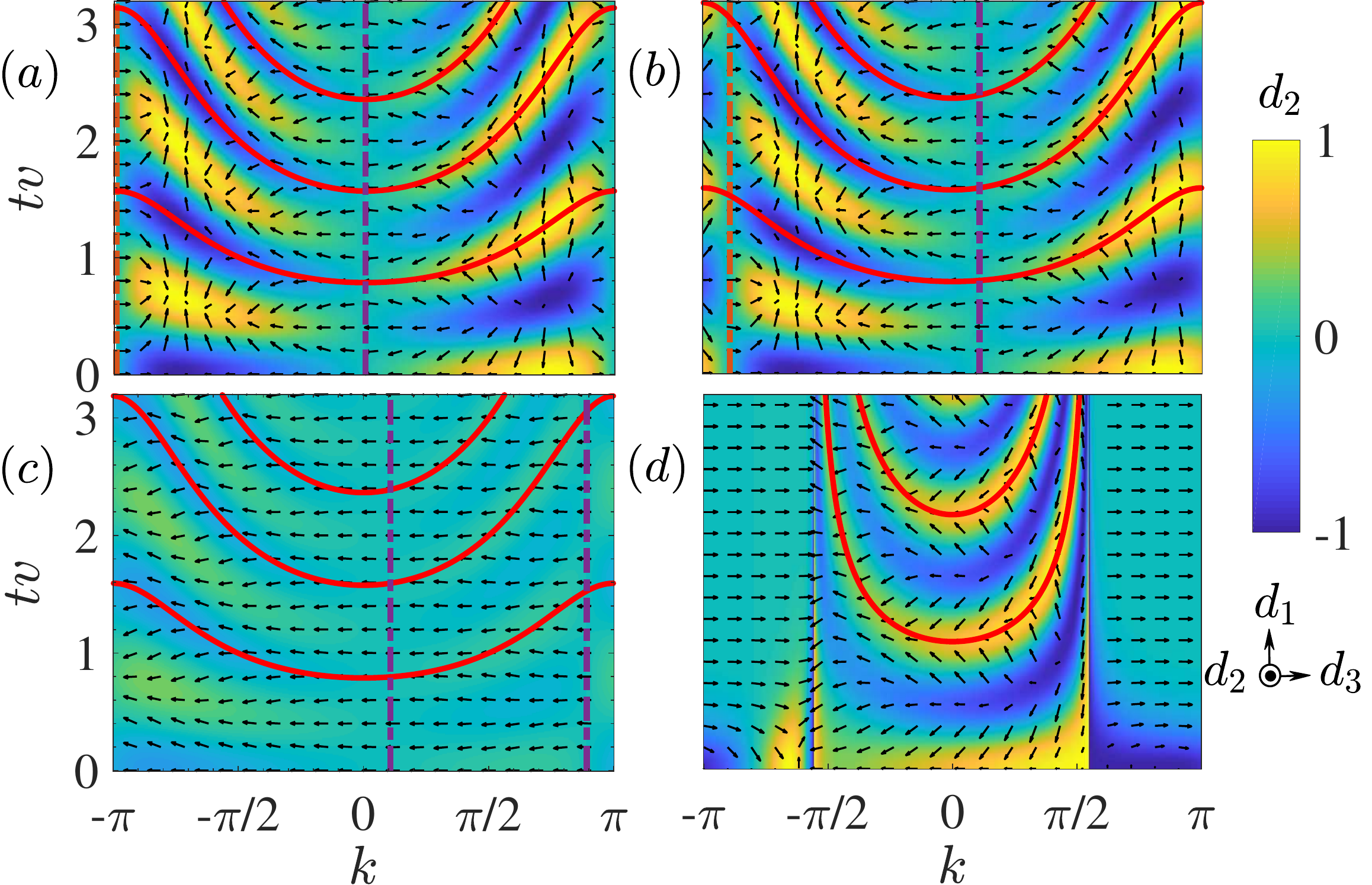}
\caption{Spin texture $\mathbf{d}(k,t)$ in the $k$-$t$ space for quench processes between $H^\text{i}$ and $H^\text{f}$ characterized by : (a) $(u^\text{i}=0, w^\text{i}=v/2)$ and $(u^\text{f}=0, w^\text{f}=3v)$;
(b) $(u^\text{i}=v/3, w^\text{i}=v/2)$ and $(u^\text{f}=v/3, w^\text{f}=3v)$;
(c) $(u^\text{i}=v/3, w^\text{i}=2v)$ and $(u^\text{f}=v/3, w^\text{f}=3v)$;
(d) $(u^\text{i}=v/3, w^\text{i}=v/2)$ and $(u^\text{f}=3v, w^\text{f}=3v)$.
We fix $v^{\text{i}}=v^{\text{f}}=v$ in the quenches, while the dynamic is unitary in (a) and non-unitary in (b)(c)(d).
The vertical dash-dotted lines in (a-c) indicate fixed points with $c_+=0$ (purple) and $c_-=0$ (orange), respectively. Skyrmion lattices associated with finite dynamic Chern numbers only emerge in (a) and (b). The solid red lines in the horizontal direction mark each period $n\pi/E^{\text{f}}_k$ ($n=1,2,...$) of spin oscillation in regions with real $E^{\text{f}}_k$. The spin dynamics is non-oscillatory in regions with imaginary $E^{\text{f}}_k$, as show in (d).}
\label{fig:skyrmions}
\end{figure}

The emergence of finite dynamic Chern numbers is underpinned by skyrmion lattices in the spin texture of $\bm{d}(k,t)$~\cite{Ueda17}.
As illustrated in Fig.~\ref{fig:skyrmions}(a)(b), when the system is quenched across the topological phase boundary with  $|\nu^\text{i}-\nu^\text{f}|=1$, two fixed points of different kinds exist in the BZ. In the unitary limit [Fig.~\ref{fig:skyrmions}(a)], the fixed points are pinned at $k=0$ and $\pi$~\cite{Chen17,Ueda17}. In the more general non-unitary case [Fig.~\ref{fig:skyrmions}(b)], fixed points deviate from $0$ and $\pi$ and need to be solved from $c_\pm(k_m)=0$.
Pairs of vortices with positive (yellow) or negative (blue) vorticity emerge in the spin texture between adjacent fixed points on the plane of $d_1$-$d_3$, with vortex cores given by $d_2(k,t)=\pm 1$. These vortices can be mapped to a lattice of skyrmions, whose topological charges are essentially the dynamic Chern numbers.
In contrast, skyrmions are absent when $H^\text{i}$ and $H^\text{f}$ belong to the same topological phase [Fig.~\ref{fig:skyrmions}(c)].

For comparison, in Fig.~\ref{fig:skyrmions}(d), we plot the spin texture when $H^\text{f}$ is in the $\mathcal{PT}$-symmetry-broken regime. As expected, in the momentum range where $E^{\text{f}}_k$ is imaginary, the spin texture approaches a steady state in the long-time limit, in sharp contrast to the periodic spin dynamics in the momentum range with real $E^{\text{f}}_k$.
We note that coincidental fixed points may still exist in the momentum range where $E^{\text{f}}_k$ is real~\cite{supp}, but their number is no longer directly related to the topology of $H^\text{i}$ and $H^\text{f}$.

\begin{figure}
\includegraphics[width=8.5cm]{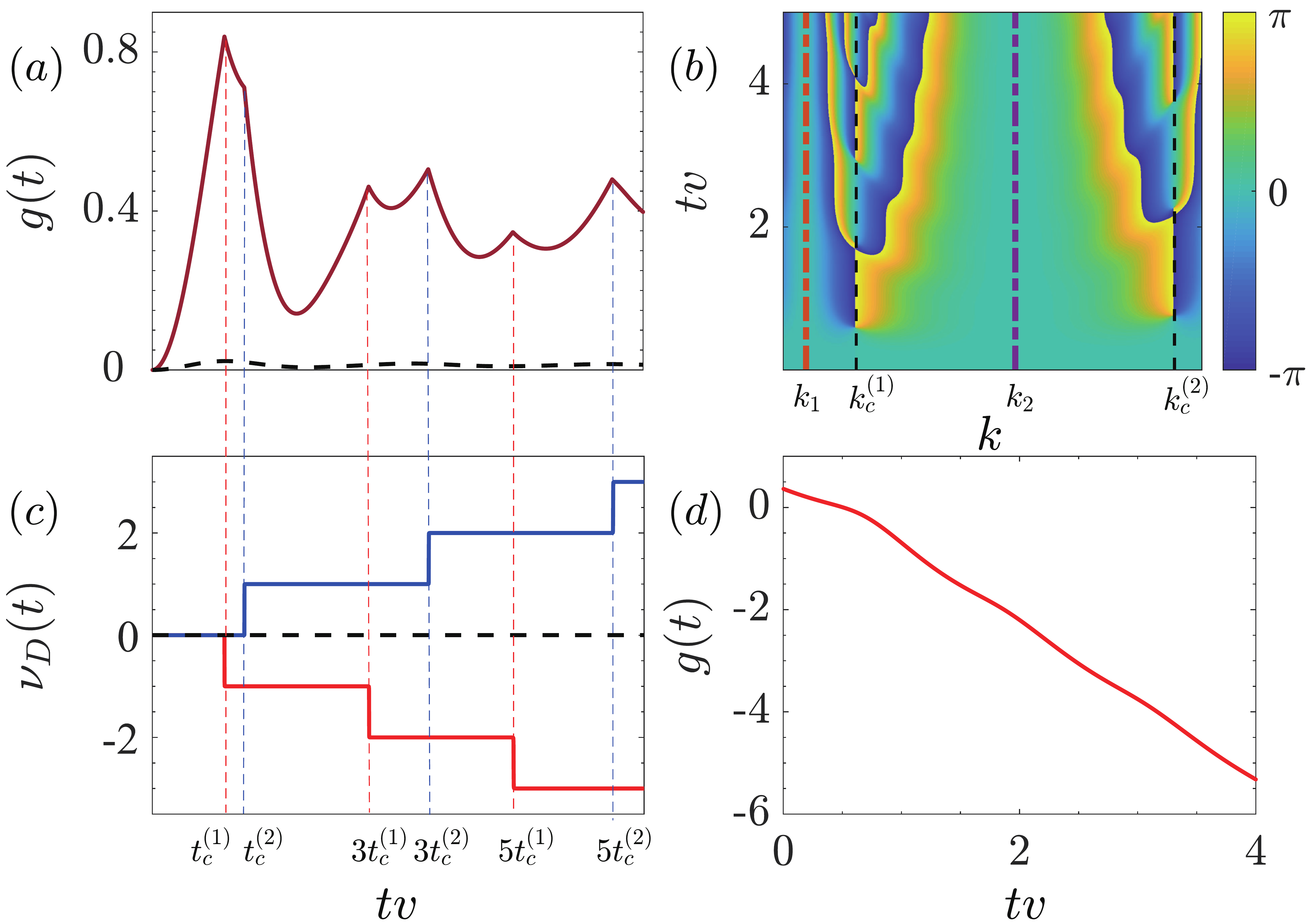}
\caption{(a) The rate function $g(t)$, (b) the PGP $\phi_k^G(t)$, and (c) the dynamic topological order parameter $\nu_D$ for the quench in Fig.~\ref{fig:skyrmions}(b). At fixed points $k_m$ ($m=1,2$), $c_-(k_1)=0$ and $c_+(k_2)=0$. $k_c^{(m)}$ and $t_c^{(m)}=\pi/(2E^{\text{f}}_{k_c^{(m)}})$ are the corresponding critical points. Two distinct types of $\nu_D(t)$ exist in (c), obtained by setting the range of integral in Eq.~(\ref{eqn:nuDf}) as $(k_1,k_2)$ (red) and $(k_2,k_1+2\pi)$ (blue), respectively.
The black dashed lines in (a) and (c) correspond to the quench in Fig.~\ref{fig:skyrmions}(c), while the quench in (d) is the same as that in Fig.~\ref{fig:skyrmions}(d).}
\label{fig:figdtop}
\end{figure}

{\it Dynamic quantum phase transition:---}
Fixed points in the non-unitary quench dynamics further give rise to DQPTs, where physical quantities become nonanalytic at critical times. Interestingly, we find that while DQPTs occur bi-periodically in time for non-unitary quench dynamics, critical points of DQPTs generically emerge as vortex cores in the momentum-time-space spin texture, which provides a crucial link between different emergent topological phenomena.

The central object in the theory of DQPT is the Loschmidt amplitude defined as the inner product of a time-evolved state with the initial state~\cite{Heyl17}. In non-unitary time evolutions, we generalize the Loschmidt amplitude as
\begin{align}
\mathcal{G}(t)=\prod_{k\in \text{1BZ}} \mathcal{G}_{k}(t)=\prod_{k\in \text{1BZ}}\left \langle\chi_k(0)|\psi_k(t)\right\rangle,\label{eqn:Losch}
\end{align}
where $\langle\chi_k(t)|$ is the associated state of $|\psi_k(t)\rangle$ defined previously. It is straightforward to derive $\mathcal{G}_{k}(t)=|c_{-}|^{2}e^{iE_{k}^{\text{f}}t}+|c_{+}|^{2}e^{-iE_{k}^{\text{f}}t}$.

DQPTs occur when the rate function $g(t)=-\frac{1}{L}\ln \left(\left|\mathcal{G}(t)\right|^2\right)$ exhibits nonanalyticities during the time evolution, which are caused by Fisher zeros~\cite{Heyl17}, where $\mathcal{G}_{k_c}(t_c)=0$ at critical points $k_c$ and $t_c$ in the dynamics. According to the expression of $\mathcal{G}_{k}(t)$, Fisher zeros, and hence DQPTs, occur periodically at $t_c=(n+\frac{1}{2})\pi/E^{\text{f}}_{k_c}$ ($n\in \mathbb{N}$) when $E_{k_c}^{\text{f}}$ is real. Here $k_c$ satisfies $|c_{-}(k_c)|=|c_{+}(k_c)|$.
When the system is quenched between different topological phases in the $\mathcal{PT}$-symmetry-preserving regime, fixed points with $c_+=0$ and those with $c_-=0$ always emerge in pairs. As $|c_+|-|c_-|$ are continuous functions of $k$, there must be at least one critical momentum satisfying $|c_+(k_c)|=|c_-(k_c)|$ inbetween two fixed points of different kinds. We thus conclude that DQPTs necessarily occur in this case.

Interestingly, at critical points, $d_2(k_c,t_c)=1$, which lie on the equator of the Bloch sphere~\cite{supp}. The critical points $(k_c,t_c)$ thus correspond to vortex cores with positive vorticity in the momentum-time spin texture [Fig.~\ref{fig:skyrmions}(a)(b)]. As vortices are manifestations of skyrmions, whose topological charges are the dynamic Chern numbers, the emergent dynamic topological phenomena are related through the spin texture.

DQPTs are characterized by the dynamic topological order parameter $\nu_D$, which is defined through the Pancharatnam geometric phase (PGP) $\phi_{k}^{G}(t)$ as
\begin{equation}
  \nu_{D}(t)=\frac{1}{2\pi}\int_{k_{m}}^{k_{n}}\frac{\partial\phi_{k}^{G}(t)}{\partial k}dk.
\label{eqn:nuDf}
\end{equation}
Here, $k_m$ and $k_n$ are fixed points of different kinds, and $\phi_{k}^{G}(t)=\phi_{k}(t)-\phi_{k}^{\text{dyn}}(t)$, where $\phi_{k}(t)$ is defined through $\mathcal{G}_{k}(t)=|\mathcal{G}_{k}(t)|e^{i \phi_{k}(t)}$ and the dynamic phase
$\phi_{k}^{\text{dyn}}(t)=-\int_{0}^{t}\langle\chi_k(t')|H_k^{\text{f}}|\psi_k(t')\rangle dt'$.
At critical points, $\mathcal{G}_{k}(t)$ vanishes, which lead to abrupt jumps in $\phi_{k}^{G}(t)$ and $\nu_D(t)$. Further, as $\phi_k^{G}(t)$ also vanishes at fixed points, $\nu_D$ characterizes the $S^1\rightarrow S^1$ mapping from the momentum submanifold between $k_m$ and $k_n$ to $e^{i\phi_k^{G}(t)}$ on the same momentum submanifold. $\nu_D(t)$ is therefore quantized despite the non-unitary time evolution.

In Fig.~\ref{fig:figdtop}(a)(b)(c), we show typical $g(t)$, $\phi_k^{G}(t)$, and $\nu_D(t)$ for a quench between different topological phases in the $\mathcal{PT}$-symmetry-preserving regime. The BZ is divided into two submanifolds by the fixed points $k_1$ and $k_2$, where each submanifold contains a critical momentum, labeled as $k_c^{(1)}$ and $k_c^{(2)}$, respectively.
These critical momenta give rise to two distinct critical time scales $t^{(1)}_c$ and $t^{(2)}_c$, and thus an apparent bi-periodicity in the occurrence of DQPTs [Fig.~\ref{fig:figdtop}(a)], in contrast to its single-period counterpart in the unitary limit ($u=0$)~\cite{Heyl15,Heyl17}.
This is due to the breaking of time reversal symmetry under the non-Hermitian SSH Hamiltonian (\ref{eqn:H}), such that the double degeneracy of Fisher zeros in the Hermitian case is lifted~\cite{Dora15}.
Correspondingly, two distinct types of $\nu_D(t)$ exist, accounting for DQPTs occurring with the period $t_c^{(1)}$ and $t^{(2)}_c$, respectively [Fig.~\ref{fig:figdtop}(c)]. In contrast, when $H^\text{i}$ and $H^\text{f}$ have the same winding numbers, $g(t)$ is a smooth function in time and $\nu_D(t)=0$ [dashed lines in Fig.~\ref{fig:figdtop}(a)(c)].

Finally, when $H^\text{f}$ is in the $\mathcal{PT}$-symmetry-broken regime, DQPT is not guaranteed~\cite{supp}, even if $H^\text{i}$ and $H^\text{f}$ possess different winding numbers. In Fig.~\ref{fig:figdtop}(d), we demonstrate the typical rate function $g(t)$ for the quench process in Fig.~\ref{fig:skyrmions}(d), where neither fixed points nor DQPTs are present.

{\it Final remarks:---}
We have shown that dynamic topological phenomena emerge in the quench dynamics of a $\mathcal{PT}$-symmetric non-Hermitian SSH model.
These dynamic topological phenomena are connected with discrete fixed points in the post-quench dynamics, and generically emerge in the non-unitary time evolution when the system is quenched between different topological phases in $\mathcal{PT}$-symmetry-preserving regime. Given the recent experimental observation of spontaneous $\mathcal{PT}$-symmetry breaking in cold atoms~\cite{leluo} and topological edge states in $\mathcal{PT}$-symmetric quantum-walk dynamics~\cite{PTsymm2}, we expect that the dynamic topological phenomena discussed here can be probed using cold atoms loaded into a superlattice with engineered on-site loss, or through discrete-time quantum-walk dynamics in photonic configurations. For either the cold atom or the photonic setup, tomography of instantaneous time-evolved states is needed to probe the dynamic topological phenomena.

\begin{acknowledgments}
\textit{Acknowledgement:--}
This work has been supported by the National Natural Science Foundation of China (Grant Nos. 11434007, 11474049, 11674056, and 11522545), the National Key Research and Development Program of China (Grant Nos. 2016YFA0301700, 2017YFA0304203), Changjiang Scholars and Innovative Research Team in University of Ministry of Education of China (Grant No. IRT13076), and the Natural Science Foundation of Jiangsu Province (Grant No. BK20160024). W. Y. acknowledges support from the ``Strategic Priority Research Program (B)" of the Chinese Academy of Sciences (Grant No. XDB01030200). X. Q. and T.-S. D. contributed equally to this work.
\end{acknowledgments}

\newpage
\begin{widetext}
\appendix

\renewcommand{\thesection}{\Alph{section}}
\renewcommand{\thefigure}{S\arabic{figure}}
\renewcommand{\thetable}{S\Roman{table}}
\setcounter{figure}{0}
\renewcommand{\theequation}{S\arabic{equation}}
\setcounter{equation}{0}

\section{Supplemental Materials}

In this Supplemental Materials, we discuss in detail the justification of topological invariants through the global Berry phase, the evolution of the density matrix, the proof for the existence of fixed points, the calculation of the dynamic Chern number, the steady-state solution for $\mathcal{PT}$-symmetry-broken final Hamiltonians, and coincidental fixed points in the $\mathcal{PT}$-symmetry-broken regime.

\section{Topological invariant through the global Berry phase}

Topological invariants defined through the global Berry phase $\varphi_B$ have the advantage that they invoke the formalism of biorthonormal basis and characterize topological properties in both the $\mathcal{PT}$-symmetry-preserving and broken regimes. The resulting winding number $\nu=\varphi_B/2\pi$ has a geometric interpretation, i.e., the number of times the unit vector $\bm{n}=\frac{1}{\sqrt{h^2_1+h^2_2}}(h_1,h_2,0)$, which lies in the $x$-$y$ plane, winds around the $z$-axis as $k$ varies through the 1BZ. Here, $\{x,y,z\}$ corresponds to the indices $\{1,2,3\}$, respectively. This geometric interpretation thus makes connection with previously defined generalized winding numbers in non-Hermitian settings~\cite{RLL16,Zeunerprl,pxprl}. It also reduces to the generalized Zak phase for $\mathcal{PT}$-symmetric non-Hermitian systems in the $\mathcal{PT}$-symmetry-preserving regime~\cite{KMKO16,PTsymm2}.

To explicitly demonstrate that the winding number defined through the global Berry phase serves as the topological invariant of the system, we numerically calculate the eigenspectra $\widetilde{E}$ of Hamiltonian (1) in various regimes under open boundary conditions. As illustrated in Fig.~\ref{fig:edgestates}, localized topological edge states with $\text{Re}(\widetilde{E})=0$ emerge in both $\mathcal{PT}$-symmetry-preserving and broken regimes, so long as $\nu=1$. In the $\mathcal{PT}$-symmetry-broken case, bulk states with $\text{Re}(\widetilde{E})=0$ also exist, but their wave functions are extended and the imaginary parts of their eigenenergies are smaller than that of the edge state.

\section{Evolution of the density matrix}

The time-dependent density matrix in the $k$-sector is written as
\begin{align}
\rho(k,t)=\frac{1}{2}\left[\tau_0+\bm{d}(k,t)\cdot\boldsymbol{\tau}\right].
\label{Eq:Sdensitymatrix}
\end{align}

When $E_k^\text{f}$ is real, we have
\begin{align}
d_0&=c^*_+ c_+ + c^*_- c_-,\\
d_1&=(c^*_-c_+e^{-i2E_k^{\text{f}}t}+\text{c.c.})/d_0,\\
d_2&= i(c^*_-c_+e^{-i2E_k^{\text{f}}t}-\text{c.c.})/d_0,\\
d_3&=(c^*_+ c_+ - c^*_- c_-)/d_0,
\end{align}
where $c_\mu=\langle\chi^{\text{f}}_\mu|\psi^{\text{i}}_-\rangle$ as defined in the main text. Apparently, while $\bm{d}(k,t)$ is a real unit vector, $d_3$ is time independent, and both $d_1$ and $d_2$ are oscillatory in time with a momentum-dependent period $t_0=\pi/E^\text{f}_k$. The fixed-point condition is satisfied when $c_-=0$ or $c_+=0$, as $d_1=d_2=0$ and $|d_3|=1$. We further parameterize $\bm{d}(k,t)$ as follows
\begin{align}
d_1(\theta,\varphi)&=\sin\theta\cos(\varphi+\delta\varphi),\\
d_2(\theta,\varphi)&=\sin\theta\sin(\varphi+\delta\varphi),\\
d_3(\theta)&=\cos\theta,
\end{align}
where the parameters $(\theta,\varphi,\delta\varphi)$ are defined through
\begin{align}
c_+&=\sqrt{d_0}e^{i\varphi_+}\cos\frac{\theta}{2},\\
c_-&=\sqrt{d_0}e^{i\varphi_-}\sin\frac{\theta}{2},\\
\varphi&=2E^{\text{f}}_k t,\\
\delta\varphi &=\varphi_- - \varphi_+.
\end{align}
Here, $\theta\in [0,\pi]$, $\varphi\in [0,2\pi)$, and $\delta\varphi$ is a constant. $\bm{d}(k,t)$ therefore depicts a vector evolving on the Bloch sphere characterized by $(\theta,\varphi)$, as illustrated in Fig.~2(a) of the main text.
We identify $\bm{d}(k,t)$ associated with the fixed point $c_-=0$ ($c_+=0$) as the north (south) pole of the Bloch sphere. Hence, when $|c_-|=|c_+|$, $d_3=0$ and the corresponding $\bm{d}(k,t)$ lies on the equator of the Bloch sphere.

\begin{figure}
\includegraphics[width=15cm]{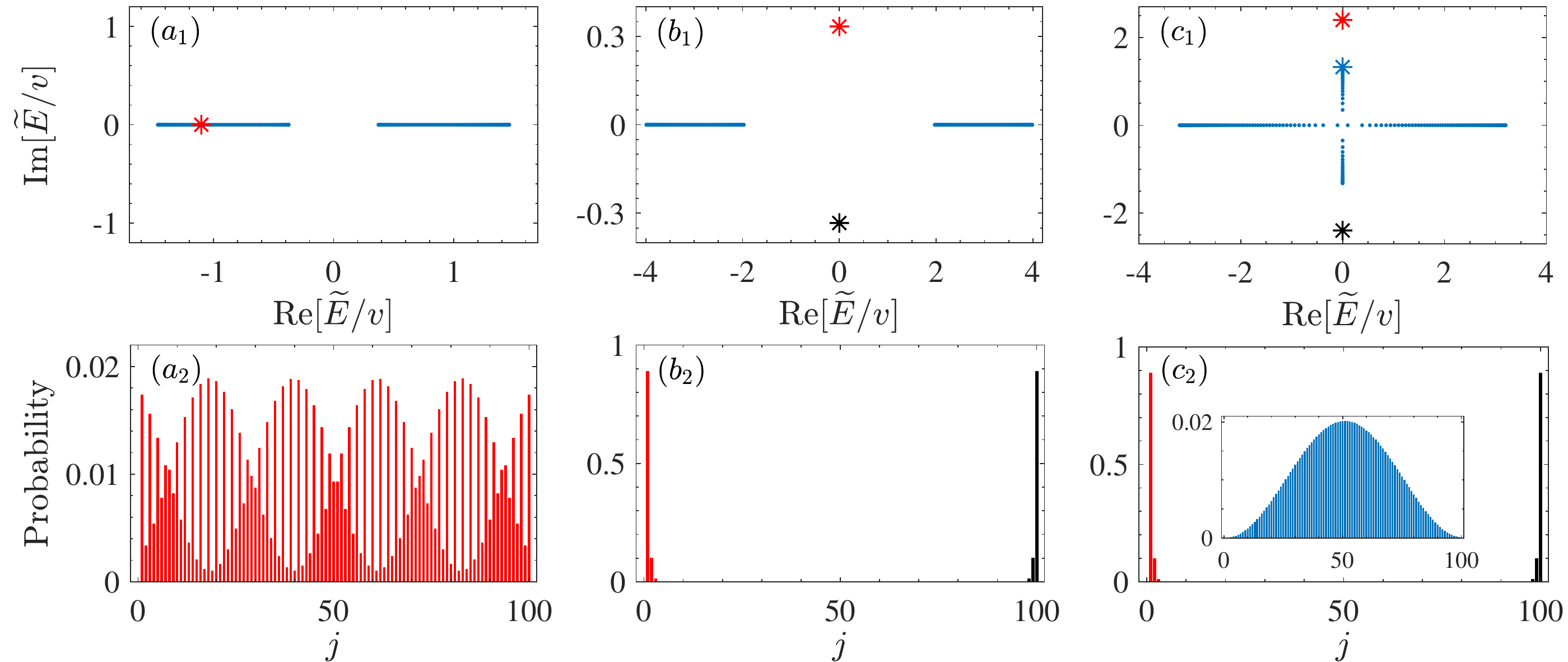}
\caption{Eigenspectra and eigen wavefunctions for non-Hermitian SSH models with open boundary conditions. We numerically diagonalize the Hamiltonian (1) with open boundary conditions on a lattice with $L=100$ sites. Eigenspectra are shown in the first row, and eigen wave functions corresponding to states marked by asterisks are shown in the second row. The parameters are:
(a1)(a2) ($u=1/3v,w=1/2v$) and $\nu=0$; (b1)(b2) ($u=1/3v,w=3v$) and $\nu=1$; (c1)(c2) ($u=2.4v,w=3v$) and $\nu=1$. (c1)(c2) are in the $\mathcal{PT}$-symmetry-broken regime. While a typical $\mathcal{PT}$-symmetry-preserving bulk state with an extended wave function is shown in (a2), topological edge states localized at two ends of the lattice emerge in both the $\mathcal{PT}$-symmetry-preserving regime (b1)(b2) and the $\mathcal{PT}$-symmetry-broken regime (c1)(c2), where $\nu=1$. Note that $\mathcal{PT}$-symmetry-broken bulk states with imaginary eigenenergies are pressent in the $\mathcal{PT}$-symmetry-broken regime (c1)(c2), but their spatial wave functions are extended rather than localized, as illustrated in the inset of (c2).}
\label{fig:edgestates}
\end{figure}

When $E^{\text{f}}_k$ is imaginary, we have
\begin{align}
d_0&=c^*_+ c_+e^{-i2E^\text{f}_kt} + c^*_- c_-e^{i2E^\text{f}_kt},\\
d_1&=(c^*_-c_++c.c.)/d_0,\\
d_2&= i(c^*_-c_+-c.c.)/d_0,\\
d_3&= (c^*_+ c_+e^{-i2E^\text{f}_kt} - c^*_- c_-e^{i2E^{\text{f}}_kt})/d_0.
\end{align}
Without loss of generality, we have assume $\text{Im}(E^{\text{f}}_k)>0$.
We parameterize $\mathbf{d}(k,t)$ as
\begin{align}
d_1(\theta,\varphi)&=\sin\theta\cos\varphi,\\
d_2(\theta,\varphi)&=\sin\theta\sin\varphi,\\
d_3(\theta,\varphi)&=\cos\theta,
\end{align}
where the parameters $(\theta,\varphi)$ are defined through
\begin{align}
c_+=&\sqrt{d_0}e^{iE^\text{f}_kt}e^{i\varphi_+}\cos\frac{\theta}{2},\\
c_-=&\sqrt{d_0}e^{-iE^\text{f}_kt}e^{i\varphi_-}\sin\frac{\theta}{2},\\
\varphi=&\varphi_--\varphi_+.
\end{align}
Here, $\theta\in[0,\pi]$, $\varphi\in[0,2\pi)$. In this case, $\bm{d}(k,t)$ is still a real unit vector on a Bloch sphere characterized by $(\theta,\varphi)$. As we show in the next section, one can prove $|c_+|\equiv |c_-|$ in this case. Therefore, we have $d_3(k,t=0)=0$, $\displaystyle\lim_{t\to\infty}d_3(k,t)=1$, and $\displaystyle\lim_{t\to\infty}d_1(k,t)=\displaystyle\lim_{t\to\infty}d_2(k,t)=0$. The vector $\bm{d}(k,t)$ thus necessarily lies on the equator of the Bloch sphere at $t=0$ and approaches the north pole in the long-time limit, as illustrated in Fig.~2(b) of the main text.
The density matrix in the corresponding $k$-sector approaches a steady-state value in the long-time limit, and fixed points do not exist.

\begin{figure}
\includegraphics[width=15cm]{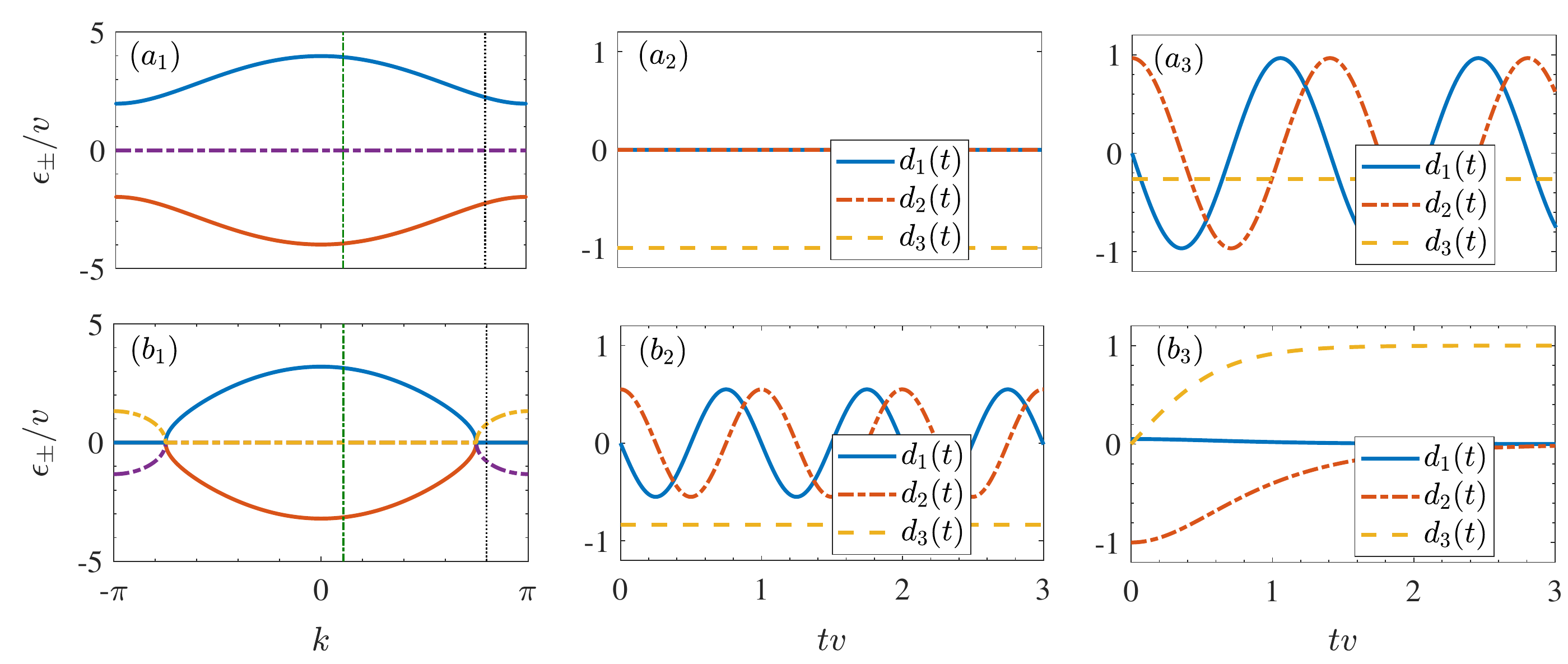}
\caption{(a1) Eigenenergy spectra for $H^\text{f}_a$ characterized by ($u^\text{f}=v/3,v^\text{f}=v,w^\text{f}=3v$). (a2)(a3) Time evolution of $\bm{d}(k,t)$ at fixed momenta as the system is quenched into $H^\text{f}_a$. (b1) Eigenenergy spectra for $H^\text{f}_b$ characterized by($u^\text{f}=2.4v,v^\text{f}=v,w^\text{f}=3v$). (b2)(b3) Time evolution of $\bm{d}(k,t)$ at fixed momenta as the system is quenched into $H^\text{f}_b$.
For the time evolution, the momenta are taken at the fixed point $k\approx 0.34$ in (a2)(b2), and at $k=4\pi/5$ in (a3)(b3), as indicated by the vertical green and black lines in (a1)(b1). In all quench processes, the initial state in each $k$-sector is the eigenstate $|\psi^{\text{i}}_-\rangle$ of the initial Hamiltonian $H^\text{i}_k$, corresponding to ($u^\text{i}=v/3,v^\text{i}=v,w^\text{i}=v/2$).}
\label{fig:dispersion}
\end{figure}


In Fig.~\ref{fig:dispersion}, we demonstrate typical time evolutions of $\bm{d}(k,t)$ at different momenta, as the system is quenched from an initial Hamiltonian in the  $\mathcal{PT}$-symmetry-preserving regime to final Hamiltonians in either the $\mathcal{PT}$-symmetry-preserving regime or the symmetry-broken regime, respectively. At fixed points [Fig.~\ref{fig:dispersion}(a2)], $\bm{d}(k,t)$ is time independent. At momenta with real $E^{\text{f}}_k$ [Fig.~\ref{fig:dispersion}(a3)(b2)], $\bm{d}(k,t)$ is oscillatory, regardless of whether the final Hamiltonian is $\mathcal{PT}$-symmetry-preserving or not. At momenta with imaginary $E^{\text{f}}_k$ when the final Hamiltonian is in the $\mathcal{PT}$-symmetry-broken regime [Fig.~\ref{fig:dispersion}(b3)], $\bm{d}(k,t)$ approaches a steady-state value in the long-time limit.

\section{Existence of fixed points}

In this section, we show that the existence and number of fixed points are intimately connected with the generalized winding numbers of $H^\text{i}$ and $H^\text{f}$, when both Hamiltonians belong to the $\mathcal{PT}$-symmetry-preserving regime with completely real eigenspectra.
First, we parameterize the initial state and the left eigenstate of the final Hamiltonian at momentum $k$, respectively, as $|\psi^\text{i}_-\rangle=\frac{1}{\sqrt{2\cos2\Omega^\text{i}}}
\left(-e^{-i\Omega^\text{i}},e^{i\phi^\text{i}}e^{i\Omega^\text{i}}\right)^T$ and $\langle\chi^{\text{f}}_{\pm}|=\frac{1}{\sqrt{2\cos2\Omega^{\text{f}}}}
\left(\pm e^{\pm i\Omega^{\text{f}}},e^{-i\phi^{\text{f}}}e^{\mp i\Omega^{\text{f}}}\right)$. Here, $\sin2\Omega^\beta = u^\beta/h^\beta$ ($\beta=\text{i},\text{f}$), $h^\beta e^{i\phi^\beta}=h^\beta_1+ih^\beta_2$, where components of $\bm{h}^{\text{i}}$ and $\bm{h}^{\text{f}}$ are associated with $H^{\text{i}}_k$ and $H^{\text{f}}_k$, respectively. Note that $\Omega^{\beta}$ is real when $E^{\beta}_k$ is real.
Importantly, $\phi^{\beta}$ is the polar angle of the vector $\bm{h}^{\beta}$, which is associated with the generalized winding number $\nu^{\beta}$ of the Hamiltonian $H^{\beta}$ through $\nu^{\beta}=\oint dk \partial \phi^{\beta}/\partial k$ ($k\in \text{1BZ}$).

We then have
\begin{align}
c_\pm=\frac{\mp e^{-i(\Omega^i\mp\Omega^{\text{f}})}+e^{i\phi^{0}}e^{i(\Omega^{\text{i}}\mp\Omega^{\text{f}})}}
{2\sqrt{\cos2\Omega^{\text{i}}\cos2\Omega^{\text{f}}}},\label{eqn:Cpm}
\end{align}
where $\phi^0=\phi^{\text{i}}-\phi^{\text{f}}$. Consider a unit vector $\bm{n}^0$ on the $x$-$y$ plane whose polar angle is given by $\phi^0$. As $k\in \text{1BZ}$, the number of times $\bm{n}^0$ winds around the $z$-axis is therefore $\oint dk \partial \phi^{0}/\partial k=\nu^{\text{i}}-\nu^{\text{f}}$.
From Eq.~(\ref{eqn:Cpm}), the condition for $c_\pm=0$ is $\phi^0=2(\Omega^\text{f}-\Omega^\text{i})$ (mod $2\pi$) and $\phi^0=\pi-2(\Omega^\text{f}+\Omega^\text{i})$ (mod $2\pi$), respectively. Therefore, the number of fixed points with $c_+=0$ or $c_-=0$ should be at least $\left|\nu^{\text{i}}-\nu^{\text{f}}\right|$ each.

Conversely, when $H^\text{f}$ is in the $\mathcal{PT}$-symmetry-broken regime, the corresponding $E^{\text{f}}_k$ becomes imaginary for a certain range of $k$. At these momenta, $\Omega^{\text{f}}$ becomes complex, which invalidates the argument following Eq.~(\ref{eqn:Cpm}). Further,
we have
\begin{align}
|c_\pm(k)|^2=\frac{\cosh2\omega^\text{f}_k-\sin(\phi^0_k+2\Omega^\text{i}_k)}
{2\cos2\Omega^\text{i}_k\sinh2\omega^\text{f}_k},
\end{align}
where $\omega_k^\text{f}=\text{Im}(\Omega^\text{f}_k)$. Hence $|c_+(k)|\equiv|c_-(k)|$ in this case. Note that $c_\pm$ cannot vanish simultaneously at the same $k$, as otherwise $|\psi^\text{i}_-(k)\rangle=\sum_\mu c_\mu(k)|\psi^\text{f}_\mu(k)\rangle=0$.

When the system is quenched between different topological phases in the $\mathcal{PT}$-symmetry-preserving regime, fixed points with $c_+=0$ and those with $c_-=0$ always emerge in pairs. As we have discussed in the main text, this gives rise to dynamic Chern numbers and emergent skyrmions in the momentum-time space. As $|c_+|-|c_-|$ is a continuous function of $k$, there must be at least one critical momentum satisfying $|c_+(k_c)|=|c_-(k_c)|$ inbetween two fixed points of different kinds. We thus conclude that DQPTs also occur in this case.

\section{Dynamic Chern number}

Between two arbitrary fixed points $k_m$ and $k_n$, the dynamic Chern number is defined as
\begin{equation}
C_{mn} = \frac{1}{4\pi}\int_{k_m}^{k_{n}}dk\int_{0}^{t_0}dt[\bm{d}(k,t)\times\partial_t \bm{d}(k,t)]\cdot\partial_k \bm{d}(k,t).
\end{equation}
The Chern number is well-defined so long as $E^{\text{f}}_k$ is real in the momentum submanifold spanned by $k_m$ and $k_n$.

\begin{figure}
\includegraphics[width=13cm]{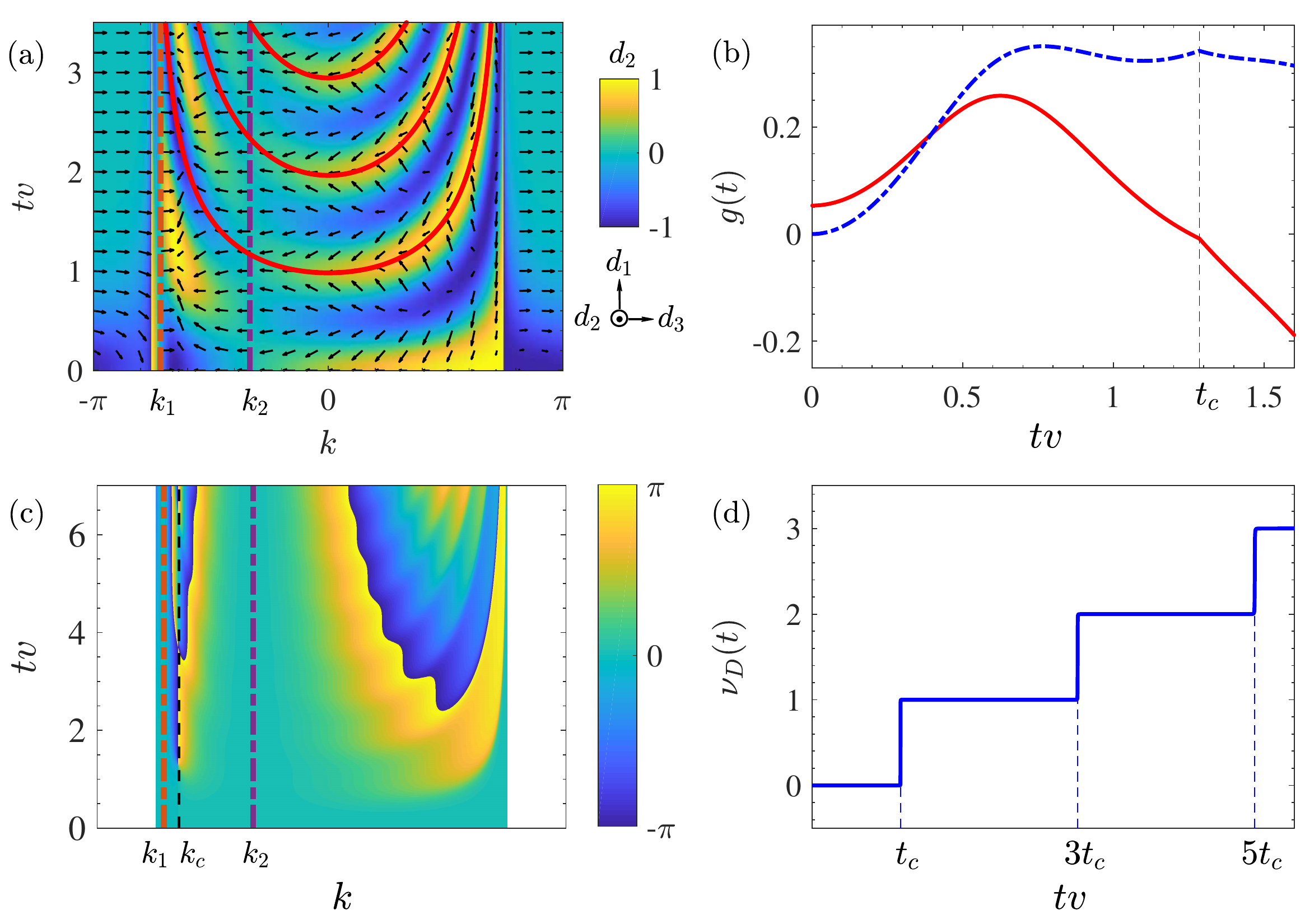}
\caption{Coincidental topological phenomena when the final Hamiltonianis in the $\mathcal{PT}$-symmetry-broken regime. The initial Hamiltonian is characterized by $(u^{\text{i}}=v/3,v^{\text{i}}=v,w^{\text{i}}=v/2)$, and the final Hamiltonian is characterized by $(u^{\text{f}}=2.4v,v^{\text{f}}=v,w^{\text{f}}=3v)$. (a) Spin texture $\bm{d}(k,t)$ in the $k$-$t$ plane. (b) Rate function $g(t)$ of the quench process (red solid), where contribution from all momenta in $\text{1BZ}$ are considered.
For comparison, we also plot the rate function where only Loschmidt amplitude in the momentum range with real $E^{\text{f}}_k$ are considered (blue dashed). The signal for DQPT is more apparent in the latter.
(c) PGP in the momentum range with real $E^{\text{f}}_k$. Note that PGP in the momentum range with imaginary $E^{\text{f}}_k$ is not well-defined. (d) $\nu_D$ as integrated between the fixed points, where $t_c=\pi/E^\text{f}_{k_c}$}.
\label{fig:bro}
\end{figure}

Following the parameterization of $\bm{d}(k,t)$ in Eqs.~(S6-S8), we have $[\bm{d}(\theta,\varphi)\times\partial_\varphi \bm{d}(\theta,\varphi)]\cdot\partial_\theta \bm{d}(\theta,\varphi)=-\sin\theta$. This allows us to
evaluate the dynamic Chern number between two adjacent fixed poits $k_m$ and $k_{m+1}$ in the following cases:

{\it Case I: $c_+(k_{m+1})=0$ and $c_-(k_{m})=0$.---}
As there are no other fixed points inbetween $k_m$ and $k_{m+1}$, $\theta$ should be integrated in the range $[0,\pi]$. We have
\begin{equation}
C_{m,m+1}= \frac{1}{4\pi}\int_{0}^{\pi}d\theta\int_{0}^{2\pi}d\varphi[\bm{d}(\theta,\varphi)\times\partial_\varphi \bm{d}(\theta,\varphi)]\cdot\partial_\theta \bm{d}(\theta,\varphi)= \frac{1}{4\pi}\int_{0}^{\pi}d\theta\int_{0}^{2\pi}d\varphi(-\sin\theta)=-1.
\end{equation}

{\it Case II: $c_+(k_{m})=0$ and $c_-(k_{m+1})=0$.---}
Similarly, we have
\begin{equation}
C_{m,m+1} = \frac{1}{4\pi}\int_{\pi}^{0}d\theta\int_{0}^{2\pi}d\varphi[\bm{d}(\theta,\varphi)\times\partial_\varphi \bm{d}(\theta,\varphi)]\cdot\partial_\theta \bm{d}(\theta,\varphi)= -\frac{1}{4\pi}\int_{0}^{\pi}d\theta\int_{0}^{2\pi}d\varphi(-\sin\theta)=1.
\end{equation}

{\it Case III: $c_+(k_{m})=0$ and $c_+(k_{m+1})=0$, or $c_-(k_{m})=0$ and $c_-(k_{m+1})=0$.---}
In this case, $\theta$ is integrated from $0$ to $0$ or from $\pi$ to $\pi$, therefore $C_{m,m+1}=0$.

For arbitrary fixed points $k_m$ and $k_n$, by successively applying the results above between adjacent fixed points,
we have: $C_{mn}=1$ when $c_+(k_{m})=0$ and $c_-(k_{n})=0$; $C_{mn}=-1$ when $c_-(k_{m})=0$ and $c_+(k_{n})=0$; when the two fixed points are of the same kind, $C_{mn}$ vanishes.

\section{Coincidental fixed points and topological phenomena in the $\mathcal{PT}$-symmetry-broken regime}

As we have seen in Fig.~\ref{fig:dispersion}(b1), momentum regime with real $E_k$ exists even when the overall Hamiltonian $H=\sum_{k\in 1\text{BZ}}H_k$ is in the $\mathcal{PT}$-symmetry-broken regime. Hence, coincidental fixed points can still exist when $H^\text{f}$ is in the $\mathcal{PT}$-symmetry-broken regime, provided that $E^{\text{f}}_{k}$ is real at these fixed points and that the initial and final Hamiltonians belong to different topological phases.

As an example, in Fig.~\ref{fig:bro}, we illustrate the emergence of coincidental dynamic topological phenomena when $H^\text{f}$ is in the $\mathcal{PT}$-symmetry-broken regime.
As indicated by vertical dashed lines in Fig.~\ref{fig:bro}(a), two fixed points exist at $k_1\approx-2.25, k_2\approx -1.05$. As both $E^{\text{f}}_{k_1}$ and $E^{\text{f}}_{k_2}$ are real, dynamic Chern number is well-defined in the $S^2$ manifold spanned by $(k_1,k_2)$ in momentum space and $(0,t_0)$ in temporal space.

On the other hand, the existence of coincidental fixed points in the $\mathcal{PT}$-symmetry-broken regime still gives rise to critical point $k_c\approx -2.05$ in momentum space, where the $\phi^G_k(t)$ undergo abrupt changes. Correspondingly, periodic nonanalyticities exist in $g(t)$. However, as illustrated in Fig.~\ref{fig:bro}(b), the signals for DQPT get drowned out by contributions from the momentum-range with imaginary eigenenergies, whose signals get exponentially enhanced over time. In this case, quantized dynamic topological order parameter $\nu_D$ can be defined and calculated [see Fig.~\ref{fig:bro}(d)] in the momentum submanifold spanned by $k_1$ and $k_2$, with
$\nu_{D}(t)=\frac{1}{2\pi}\int_{k_1}^{k_2}\frac{\partial\phi_{k}^{G}(t)}{\partial k}dk$.

%
%
%
%
%
%

\end{widetext}


\begin{thebibliography}{99}
\bibitem{HKrmp10} M. Z. Hasan and C. L.  Kane, Rev. Mod. Phys. {\bf 82}, 3045 (2010).
\bibitem{QZrmp11} X. L. Qi and S. C. Zhang, Rev. Mod. Phys. {\bf 83}, 1057 (2011).

\bibitem{Demler10} T. Kitagawa, E. Berg, M. Rudner, and E. Demler, Phys. Rev. B {\bf 82}, 235114 (2010).
\bibitem{Zoller11} L. Jiang, T. Kitagawa, J. Alicea, A. R. Akhmerov, D. Pekker, G. Refael, J. I. Cirac, E. Demler, M. D. Lukin, and P. Zoller, Phys. Rev. Lett. {\bf 106}, 220402 (2011).
\bibitem{Zollerdiss} S. Diehl, E. Rico, M. A. Baranov, and P. Zoller, Nat. Phys. {\bf 7}, 971 (2011).
\bibitem{Levin13} M. S. Rudner, N. H. Lindner, E. Berg, and M. Levin, Phys. Rev. X {\bf 3}, 031005 (2013).
\bibitem{Gogolin15} J. Eisert, M. Friesdorf, and C. Gogolin, Nat. Phys. {\bf 11}, 124 (2015).
\bibitem{Bhaseen15} M. D. Caio, N. R. Cooper, and M. J. Bhaseen, Phys. Rev. Lett. {\bf 115}, 236403 (2015).
\bibitem{rigol} L. D'Alessio and M. Rigol, Nat. Commun. {\bf 6}, 8336 (2015).
\bibitem{Lindner16} P. Titum, E. Berg, M. S. Rudner, G. Refael, and N. H. Lindner, Phys. Rev. X {\bf 6}, 021013 (2016).
\bibitem{Vishwanath16} A. C. Potter, T. Morimoto, and A. Vishwanath, Phys. Rev. X {\bf 6}, 041001 (2016).
\bibitem{Sondhi16} V. Khemani, A. Lazarides, R. Moessner, and S. L. Sondhi, Phys. Rev. Lett. {\bf 116}, 250401 (2016).
\bibitem{Heyl13} M. Heyl, A. Polkovnikov, and S. Kehrein, Phys. Rev. Lett. {\bf 110}, 135704 (2013).
\bibitem{Heyl15} M. Heyl, Phys. Rev. Lett. {\bf 115}, 140602 (2015).
\bibitem{Heyl17} M. Heyl, arXiv:1709.07461.
\bibitem{Dora15} S. Vajna and B. Dora, Phys. Rev. B {\bf 91}, 155127 (2015).
\bibitem{BH16} J. C. Budich and M. Heyl, Phys. Rev. B {\bf 93}, 085416 (2016).
\bibitem{Balatsky} Z. Huang and A. V. Balatsky, Phys. Rev. Lett. {\bf 117}, 086802 (2016).
\bibitem{Budich16} Y. Hu, P. Zoller, and J. C. Budich, Phys. Rev. Lett. {\bf 117}, 126803 (2016).
\bibitem{Refael16} J. H. Wilson, J. C. W. Song, and G. Refael, Phys. Rev. Lett. {\bf 117}, 235302 (2016).
\bibitem{Bhaseen16} M. D. Caio, N. R. Cooper, and M. J. Bhaseen, Phys. Rev. B {\bf 94}, 155104 (2016).
\bibitem{Sondhi17} R. Moessner and S. L. Sondhi, Nat. Phys. {\bf 13}, 44 (2017).
\bibitem{Zhai17} C. Wang, P. Zhang, X. Chen, J. Yu, and H. Zhai, Phys. Rev. Lett. {\bf 118}, 185701 (2017).
\bibitem{Chen17} C. Yang, L. Li, and S. Chen, Phys. Rev. B {\bf 97}, 060304(R) (2018).
\bibitem{Ueda17} Z. Gong and M. Ueda, arXiv:1710.05289.
\bibitem{Chang18} P.-Y. Chang, Phys. Rev. B {\bf 97}, 224304 (2018).
\bibitem{Xiong-Jun1802} L. Zhang, L. Zhang, S. Niu, X. -J. Liu, arXiv: 1802.10061.
\bibitem{Xiong-Jun1807} L. Zhang, L. Zhang, X. -J. Liu, arXiv: 1807.10782.





\bibitem{RL09} M. S. Rudner and L. S. Levitov, Phys. Rev. Lett. {\bf 102}, 065703 (2009).
\bibitem{ESHK11} K. Esaki, M. Sato, K. Hasebe, and M. Kohmoto, Phys. Rev. B {\bf 84}, 205128 (2011).
\bibitem{LH13} S.-D. Liang and G.-Y. Huang, Phys. Rev. A {\bf 87}, 012118 (2013).
\bibitem{schomerus2013} H. Schomerus, Opt. Lett. {\bf 38}, 1912 (2013).
\bibitem{KMKO16} D. Kim, K. Mochizuki, N. Kawakami, and H. Obuse, arXiv: 1609.09650.
\bibitem{UedaPT} K. Kawabata, Y. Ashida, H. Katsura, and M. Ueda, arXiv:1801.03224.
\bibitem{RLL16} M. S. Rudner, M. Levin, and L. S. Levitov, arXiv: 1605.07652.
\bibitem{Gong17} L. Zhou, Q. Wang, H. Wang, and J. Gong, arXiv: 1711.10741.
\bibitem{Lieu18} S. Lieu, Phys. Rev. B {\bf 97}, 045106 (2018).
\bibitem{UedaNH} Z. Gong, Y. Ashida, K. Kawabata, K. Takasan, S. Higashikawa, and M. Ueda, arXiv:1802.07964.
\bibitem{Wang1803} S. Yao, and Z. Wang, arXiv:1803.01876.
\bibitem{Kunst18} F. K. Kunst, E. Edvardsson, J. Budich, and E. J. Bergholtz, Phys. Rev. Lett. {\bf 121}, 026808 (2018).



\bibitem{ETHcoldatom14} G. Jotzu, M. Messer, R. Desbuquois, M. Lebrat, T. Uehlinger, D. Greif, and T. Esslinger, Nature {\bf 515}, 237 (2014).
\bibitem{Weitenberg2016} N. Fl\"{a}schner, B. S. Rem, M. Tarnowski, D. Vogel, D.-S. L\"{u}hmann, K. Sengstock, and C. Weitenberg, Science {\bf 352}, 1091 (2016).
\bibitem{Jo17} B. Song, L. Zhang, C. He, T. F. J. Poon, E. Haiiyev, S. Zhang, X.-J. Liu, and G.-B. Jo, Science Advances {\bf 4}, 4748 (2018).
\bibitem{Weitenberg1709} M. Tarnowski, F. Nur-Unal, N. Flaschner, B. S. Rem, A. Eckard, K. Sengstock, and C. Weitenberg, arXiv:1709.01046.
\bibitem{Weitenberg17} N. Fl\"{a}schner, D. Vogel, M. Tarnowski, B. S. Rem, D.-S. L\"{u}hmann, M. Heyl, J. C. Budich, L. Mathey, K. Sengstock, and C. Weitenberg, Nat. Phys. {\bf 14}, 265 (2018)
\bibitem{Bellecnc} C. Poli, M. Bellec, U. Kuhl, F. Mortessagne, and  H. Schomerus, Nat. Commun. {\bf 6}, 6710 (2015).
\bibitem{Weimannnm} S. Weimann, M. Kremer, Y. Plotnik, Y. Lumer, S. Nolte, K. G. Makris, M. Segev, M. C. Rechtsman, and A. Szameit, Nat. Mater. {\bf 16}, 433 (2017).
\bibitem{PTsymm2} L. Xiao, X. Zhan, Z. H. Bian, K. K.Wang, X. Zhang, X. P. Wang, J. Li, K. Mochizuki, D. Kim, N. Kawakami, W. Yi, H. Obuse, B. C. Sanders, and P. Xue, Nat. Phys. {\bf 13}, 1117 (2017).
\bibitem{Zeunerprl} J. M. Zeuner, M. C. Rechtsman, Y. Plotnik, Y. Lumer, S. Nolte, M. S. Rudner, M. Segev, and A. Szameit, Phys. Rev. Lett. {\bf 115}, 040402 (2015).
\bibitem{pxprl} X. Zhan, L. Xiao, Z. Bian, K. Wang, X. Qiu, B. C. Sanders, W. Yi, and P. Xue, Phys. Rev. Lett. {\bf 119}, 130501 (2017).

\bibitem{iondtop} P. Jurcevic, H. Shen, P. Hauke, C. Maier, T. Brydges, C. Hempel, B. P. Lanyon, M. Heyl, R. Blatt, and C. F. Roos, Phys. Rev. Lett. {\bf 119}, 080501 (2017).

\bibitem{BB98} C. M. Bender and S. Boettcher, Phys. Rev. Lett. {\bf 80}, 5243 (1998).
\bibitem{BBJ02} C. M. Bender, D. C. Brody, and H. F. Jones, Phys. Rev. Lett. {\bf89}, 270401 (2002).
\bibitem{B07} C. M. Bender, Rep. Prog. Phys. {\bf70}, 947 (2007).

\bibitem{SSH} W. P. Su, J. R. Schrieffer, and A. J. Heeger, Phys. Rev. Lett. {\bf 42}, 1698 (1979).
\bibitem{SSHPT} B. Zhu, R. Lu, and S. Chen, Phys. Rev. A {\bf 89}, 062102 (2014).

\bibitem{GW88} J. Garrison and E. Wright, Phys. Lett. A {\bf 128}, 177 (1988).
\bibitem{supp} See Supplemental Materials.
\bibitem{DCB} D. C. Brody, J. Phys. A: Math. Theor. \textbf{47} 035305 (2014).


\bibitem{leluo} J. Li, A. K. Harter, J. Liu, L. de Melo, Y. N. Joglekar, and L. Luo, arXiv:1608.05061.

\end{thebibliography}
\end{document}